\documentclass[11pt]{article}

\usepackage{graphicx}
\usepackage[utf8]{inputenc}
\usepackage{amsthm}
\usepackage{amsmath}
\usepackage{amssymb}
\usepackage[lined,boxed]{algorithm2e}
\usepackage{geometry}     
\usepackage{amsfonts,amstext,thm-restate,xspace,url,hyperref}           
\allowdisplaybreaks

\def\E{\ensuremath{{\mathbb E}}}
\def\norm#1{\mathopen\| #1 \mathclose\|}

\newcommand{\normal}{truncated\xspace}
\newcommand{\I}{\ensuremath{\mathcal{I}}\xspace}
\newcommand{\stocload}{{\textsc{StocMakespan}}\xspace}
\newcommand{\budgstoc}{{\textsc{BudgetStocMakespan}}\xspace}
\newcommand{\GAP}{\textsc{GAP}\xspace}
\newcommand{\budgGAP}{\textsc{Budgeted GAP}\xspace}
\newcommand{\qnormsched}{\textsc{$q$-DetSched}\xspace}
\newcommand{\reward}{r}
\newcommand{\sse}{\subseteq}

\def\E{\mathbb{E}}

\def\sse{\subseteq}

\newtheorem{theorem}{Theorem}
\newtheorem{definition}{Definition}
\newtheorem{claim}{Claim}
\newtheorem{observation}{Observation}

\newtheorem{lemma}{Lemma}
\newtheorem{proposition}{Proposition}

\newcommand\ignore[1]{}

\newenvironment{pf}{
	
	\noindent{\bf Proof: }} {\hfill$\square$

}
\date{}

\title{Stochastic Load Balancing on Unrelated Machines}

\author{Anupam Gupta\thanks{Computer Science Department, Carnegie Mellon University}\and Amit Kumar\thanks{Dept. of Computer Science and Engg., IIT Delhi}\and Viswanath Nagarajan\thanks{Industrial and Operations Engineering Department, University of Michigan.} \and Xiangkun Shen$^\ddagger$}

\begin{document}
\maketitle

\begin{abstract}
We consider the problem of makespan minimization on unrelated machines when job sizes are stochastic. The goal is to find a fixed assignment of jobs to machines, to minimize the  expected value of the maximum load over all the machines. For the identical machines special case when the size of a job is the same across all machines, a constant-factor approximation algorithm has long been known. 
Our main result  is  the first constant-factor approximation algorithm for   the general case of unrelated machines. This is achieved by (i)  formulating a lower bound using an exponential-size linear program that is efficiently computable, and (ii) rounding this linear program while satisfying only a specific subset of the constraints that still suffice to bound the expected makespan. 
We also consider two generalizations. The first is the \emph{budgeted} makespan minimization problem, where the goal is to minimize the expected makespan subject to scheduling a target number (or reward) of jobs. We extend our main result to obtain a constant-factor approximation algorithm for this problem. 
The second problem involves   \emph{q-norm} objectives, where we want to minimize the expected q-norm of the machine loads.  Here we give an $O(q/\log q)$-approximation algorithm, which is a constant-factor approximation for any fixed $q$. 
\end{abstract}

\section{Introduction}
\label{sec:introduction}

We consider the problem of scheduling jobs on machines to minimize the maximum load (i.e., the problem of \emph{makespan minimization}). This is a classic NP-hard problem, with Graham's list scheduling algorithm for the identical machines being one of the earliest approximation algorithms known. If the job sizes are deterministic, the problem is fairly well-understood, with  PTASes for the identical~\cite{HochbaumS87} and related machines cases~\cite{HochbaumS88}, and a constant-factor approximation and APX-hardness~\cite{LenstraST90, ShmoysT93} for the unrelated machines case. Given we understand the basic problem well, it is natural to consider settings which are less stylized, and one step closer to modeling real-world scenarios: \emph{what can we do if there is uncertainty in the job sizes?}

We take a stochastic optimization approach where the job sizes are  random variables with known distributions. In particular,
the size of each job $j$ on machine $i$ is given by a random variable $X_{ij}$. Throughout  this paper we assume that the sizes of different jobs 
are independent of each other.  
Given just this information, an algorithm has to assign these jobs to machines, say resulting in jobs $J_i$ being assigned to each machine~$i$. The expected makespan of this assignment is 
\begin{equation}
	\E\left[ \max_{i = 1}^m \sum_{j \in J_i} X_{ij} \right], \label{eq:obj}
\end{equation}
where $m$ denotes the number of machines. The goal for the algorithm is to minimize this expected makespan. Observe that the entire assignment of jobs to machines is done up-front without knowledge of the actual outcomes of the random variables, and hence there is no adaptivity in this problem.

Such stochastic load-balancing problems are common in real-world systems where the job sizes are indeed not known, but given the large amounts of data, one can generate reasonable estimates for the distribution. Moreover static (non-adaptive) assignments are preferable in many applications as they are easier to implement. 

Inspired by work on scheduling and routing problems in several communities, Kleinberg, Rabani, and Tardos first posed the problem of approximating the expected makespan in 1997~\cite{KRT}. They gave a constant-factor approximation for the {\em identical machines} case, i.e., for the case where
for each job $j$, the sizes $X_{ij} = X_{i'j}$ for all $i, i' \in [m]$. A key concept in their result was the 
\emph{effective size} of a random variable (due to Hui~\cite{Hui}) which is a suitably scaled logarithm of the moment generating function. This effective size (denoted $\beta_m$) depended crucially on the number $m$ of machines. Roughly speaking, the algorithm in \cite{KRT} solved the {\em determinisitic} makespan minimization problem by using the  effective size $\beta_m(X_j)$ of each job $j$ as its deterministic size. The main part of their analysis involved proving that the resulting schedule also has small {\em expected} makespan when viewed as a solution to the stochastic problem.  See Section~\ref{sec:prelim} for a more detailed discussion.


Later, Goel and Indyk~\cite{GoelI99} gave better approximation ratios  for special classes of job size distributions, again for identical machines.
Despite these improvements and refined understanding of the identical machines case, the above stochastic load-balancing problem has remained open, even for the related machines setting. Recall that \emph{related	machines} refers to the  case where each machine $i$ has a \emph{speed} $s_i$, and the sizes  for each job $j$ satisfy $X_{ij}\,s_i = X_{i'j}\, s_{i'}$ for all $i,i'\in[m]$. 


\subsection{Results and Techniques}
Our main result is:
\begin{restatable}{theorem}{MMThm}
	\label{thm:main}
	There is an $O(1)$-approximation algorithm for the problem of finding
	an assignment to minimize the expected makespan  on unrelated
	machines.
\end{restatable}

Our work naturally builds on the edifice of~\cite{KRT}. However, we need several new ideas to achieve this. As mentioned above, the prior result for identical machines used the notion of effective size, which depends on the number $m$ of machines available. When  machines are not identical, consider just the ``restricted assignment'' setting where each job  needs to choose from a specific subset of machines: here it is not even clear how to define the effective size of each job. Instead of working with a single deterministic value  as the effective size of any random variable  $X_{ij}$, we use all the $\beta_k(X_{ij})$ values for $k=1,2,\cdots m$. 


Then we show that in an optimal solution, for every $k$-subset of machines, the total $\beta_k$ effective size of jobs assigned to those machines is at most some bound depending on $k$. 
Such a property for $k=m$ was also used in the algorithm for  identical machines. We then formulate a linear program (LP) relaxation that enforces such a  ``volume'' constraint  for {\em every} subset of machines. Although our LP relaxation has an exponential number of constraints, it can be solved in polynomial time using the ellipsoid algorithm and a suitable separation oracle.

Finally, given an optimal solution to this LP, we show how to carefully choose the right parameter for effective size of each job and use it to build an  approximately optimal  schedule. Although our LP relaxation has an exponential number of constraints (and it seems  difficult to preserve them all), we show that it suffices to satisfy a small subset of these constraints in the  integral solution. 
Roughly, our rounding algorithm uses the LP solution to identify the ``correct'' deterministic size for each job and then applies an existing algorithm for deterministic scheduling~\cite{ShmoysT93}.


\medskip\textbf{Budgeted Makespan Minimization.}  In this  problem,  each job $j$ has a \emph{reward} $r_j$ (having no relationship to other parameters such as its size), and we are given a target reward value $R$. The goal is to assign some subset $S \sse [n]$ of jobs whose total reward $\sum_{j \in S} r_j$ is at least $R$, and to minimize the expected makespan of this assignment. Clearly, this
generalizes the basic makespan minimization problem (by setting all rewards to one and the target $R=n$). 
\begin{restatable}{theorem}{BMMThm}
	\label{thm:main-reward}
	There is an $O(1)$-approximation algorithm for the budgeted makespan
	minimization problem on unrelated machines.
\end{restatable}
To solve this, we extend the ideas for expected makespan scheduling to include an extra constraint about  high reward. We again write a similar LP relaxation. Rounding this LP  requires some additional ideas on top of those in Theorem~\ref{thm:main}. The new ingredient is that we need to round solutions to an assignment LP with two linear constraints. To do this without violating the budget, we utilize  a ``reduction'' from the Generalized Assignment Problem to bipartite matching~\cite{ShmoysT93} as well as certain adjacency properties of the bipartite matching polytope~\cite{BR74}. 

\medskip\textbf{Minimizing $\ell_q$ Norms.} 
Finally, we consider the problem of stochastic load balancing under  $\ell_q$ norms. Note that given some assignment, we can denote the ``load'' on machine $i$ by $L_i := \sum_{j \in J_i} X_{ij}$, and the ``load vector'' by $\mathbf{L} = (L_1, L_2, \ldots, L_m)$. The expected makespan minimization problem is to minimize $\E[ \norm{ \mathbf{L} }_\infty ]$. The \emph{$q$-norm minimization problem} is the following: find an assignment of jobs to machines to minimize 
\[ \E\big[ \norm{ \mathbf{L} }_q \big] = \E \Bigg[ \bigg( \sum_{i =
	1}^m \bigg( \sum_{j \in J_i} X_{ij} \bigg)^q \bigg)^{1/q}
\Bigg]. \]
Our result for this setting is the following:
\begin{restatable}{theorem}{QNormThm}
	\label{thm:main-qnorm}
	There is an $O(\frac{q}{\log q})$-approximation algorithm for the stochastic 
	$q$-norm minimization problem on unrelated machines.
\end{restatable}
The main idea here  is to reduce this problem to a suitable instance of deterministic $q$-norm minimization with additional side constraints. We then show that existing techniques for deterministic $q$-norm minimization~\cite{AzarE05} can be extended to  obtain  a constant-factor approximation for our generalization as well. We also need to use/prove some probabilistic inequalities to relate  the deterministic sub-problem to  the stochastic problem. We note that using  general polynomial concentration inequalities~\cite{KimV00,SchudyS12} only yields an approximation ratio that is exponential in $q$. We obtain a much better  $O(q/\log q)$-approximation factor by utilizing the specific form of the norm function. Specifically, we use the  convexity of  norms, a second-moment calculation and a concentration bound~\cite{JSZ} for the $q^{th}$ moment of sums of independent random variables. 

We note that Theorem~\ref{thm:main-qnorm} implies a constant-factor approximation for any fixed $q\ge 1$. 
However, our techniques do not extend directly to provide an $O(1)$-approximation algorithms for all $q$-norms.

\subsection{Other Related Work}
\label{sec:related}

Goel and Indyk~\cite{GoelI99} considered the stochastic load balancing problem on identical machines (same setting as~\cite{KRT})  but for specific job-size distributions. For Poisson distributions they showed that Graham's algorithm achieves a $2$-approximation, and for Exponential distributions they obtained a PTAS. Kleinberg et al.~\cite{KRT} also considered stochastic versions of knapsack and bin-packing: given an overflow probability $p$, a feasible single-bin packing here corresponds to any subset of jobs such that their total size exceeds one with  probability  at most $p$. \cite{GoelI99}  gave better/simpler algorithms for these problems,  under special distributions. 

Recently, Deshpande and Li~\cite{LD11}, and Li and Yuan~\cite{LY13} considered several combinatorial optimization problems including shortest paths, minimum spanning trees, where elements have weights (which are random variables), and one would like to find a  solution (i.e. a subset of elements) whose expected utility is maximized. These results  also apply to the stochastic versions of knapsack and bin-packing from \cite{KRT} and yield  bicriteria approximations. The main technique here is a clever discretization of probability distributions. However, to the best of our knowledge,  such an approach is not  applicable to stochastic load balancing.

Stochastic scheduling  has been studied in many different contexts, by different fields (see, e.g.,~\cite{Pinedo04}). The work on approximation algorithms for these problems is more recent; see~\cite{MSU} for some early work and many references. In this paper we consider the \emph{(non-adaptive) fixed assignment model}, where jobs have to be assigned to machines up-front, and then the randomness is revealed.  Hence, there is no element of adaptivity in these problems. This makes them suitable for
settings where the decisions cannot be instantaneously implemented (e.g., for virtual circuit routing, or assigning customers to physically well-separated warehouses). A number of papers~\cite{MSU,  MegowUV06, IMP, GMUX} have considered scheduling problems in the \emph{adaptive} setting, where assignments are done online  and the assignment for a job may depend on the state of the system at the time of its assignment. See Section~\ref{sec:prelim} for a comparison of adaptive and non-adaptive settings in the load balancing problem.

Very recently (after the preliminary version of this paper appeared), Molinaro \cite{Mol18} obtained an $O(1)$-approximation algorithm for the stochastic $q$-norm problem for all $q\ge 1$, which improves over Theorem~\ref{thm:main-qnorm}. In addition to the techniques in our paper, the main idea in \cite{Mol18} is to use a different notion of effective size, based on the L-function method~\cite{Latala}. We still present our algorithm/analysis for Theorem~\ref{thm:main-qnorm} as it is conceptually simpler and may provide better constant factors for small $q$.


\section{Preliminaries}\label{sec:prelim}
The stochastic load balancing problem (\stocload) involves assigning  $n$ jobs to $m$ machines. For each job $j\in [n]$ and machine $i\in [m]$, we are given a random
variable $X_{ij}$ which denotes the processing time (size) of job $j$ on machine
$i$. We assume that the random variables $X_{ij}, X_{i',j'}$ are
independent when $j \neq j'$ (the size of job $j$ on different  machines may be correlated). We assume access to
the distribution of these random variables via some (succinct)
representation. 
A solution is a partition $\{J_i\}_{i=1}^m$ of the jobs among the machines, such that
$J_i\sse[n]$ is the subset of jobs assigned to machine $i\in[m]$. The
expected makespan of this solution is $\E  \left[ \max_{i=1}^m \, \sum_{j\in J_i} X_{ij} \right]$. 
Our goal is to find a solution which minimizes the expected makespan.  

The deterministic load balancing problem is known to be $\mathcal{NP}$-hard even on identical machines. The stochastic version introduces considerable additional complications. For example, \cite{KRT} showed that given Bernoulli random variables $\{X_j\}$, it is  $\#\mathcal{P}$-hard to compute the overflow probability, i.e. $\Pr[\sum_{j}X_j>1]$. We now show that it is $\#\mathcal{P}$-hard even to compute the objective value of a given assignment in the identical machines setting. 
\begin{theorem}
	It is $\#\mathcal{P}$-hard  to compute the expected makespan of a given assignment for stochastic load balancing on identical machines. 
\end{theorem}
\begin{pf}
	We will reduce the overflow probability problem~\cite{KRT} to this problem. Formally, an instance of the overflow probability problem is: given Bernoulli trials $Y_1,\cdots, Y_n$, where each $Y_j$ has size $s_j$ with probability $q_j$, 
	compute $\Pr\left(\sum_{j=1}^nY_j> 1\right)$.
	
	For an instance of the overflow probability problem, we construct two instances of 	stochastic load balancing with Bernoulli jobs and $m=2$ machines. For each $Y_j$ of type $(q_j,s_j)$, we create random variable $X_j$ of type $(q_j,\bar{s}_j)$ where $\bar{s}_j=Bs_j$ and $B$ is a scalar such that $\bar{s}_j\in\mathbb{Z}$, $\forall j$. We also create two additional random variable:  $Z_{B}$ of type $(1,B)$ and $Z_{B+1}$ of type $(1,B+1)$. Then the first instance contains two machines and jobs $X_1,\dots,X_n$ and $Z_B$. The second one contains two machines and jobs $X_1,\dots,X_n$ and $Z_{B+1}$. We want to compute the expected makespan of the following assignment for both instances: assign jobs $X_1,\dots,X_n$ to machine 1 and the remaining job to machine 2. We use $Obj_B$ and $Obj_{B+1}$ to denote the expected makespans  of the two instances. Then we have
	\begin{align*}
	Obj_B=&\E\left[\max\left\{B,\sum_{j=1}^nX_j\right\}\right]=B+\E\left[\max\left\{0,\sum_{j=1}^nX_j-B\right\}\right]\\
	=&B+\sum_{t\ge B+1}\Pr\left(\sum_{j=1}^nX_j\ge t\right).
	\end{align*}
	Similarly,
	\begin{align*}
	Obj_{B+1}=B+1+\sum_{t\ge B+2}\Pr\left(\sum_{j=1}^nX_j\ge t\right).
	\end{align*}
	It follows that $\Pr\left(\sum_{j=1}^nY_j> 1\right)=\Pr\left(\sum_{j=1}^nX_j\ge B+1\right)=Obj_B-Obj_{B+1}+1$. Therefore, it is $\#\mathcal{P}$-hard  to compute the expected makespan of a given assignment.
\end{pf}

\medskip
{\bf Scaling the optimal value.} Using a standard binary search approach, in order to obtain an $O(1)$-approximation algorithm for  \stocload it suffices to solve the following problem. Given a bound $M>0$, either find a solution with expected makespan $O(M)$ or establish that the optimal makespan is  $\Omega(M)$. Moreover, by scaling down all random variables by factor $M$, we may assume that   $M=1$.

We now provide some definitions and background results that will be used extensively in the rest of the paper.  

\subsection{Truncated and Exceptional Random Variables}
\label{sec:normal}

It is convenient  to divide each random variable
$X_{ij}$ into its \emph{\normal} and \emph{exceptional} parts, defined below: 
\begin{itemize}
	\item $X'_{ij} := X_{ij} \cdot \mathbf{I}_{(X_{ij}\le 1)}$ (called
	the \emph{\normal} part), and
	\item $X''_{ij} := X_{ij} \cdot \mathbf{I}_{(X_{ij}> 1)}$
	(called the \emph{exceptional} part).
\end{itemize}

The reason for doing this is that these two kinds of random variables (r.v.s) behave very differently with respect to  the expected makespan. It turns out that  expectation is a good notion of ``deterministic''  size for exceptional r.v.s, whereas one needs a more nuanced notion (called effective size) for truncated r.v.s: this is discussed in detail  below.   

We will use the following result (which follows from \cite{KRT}) to handle exceptional r.v.s. 
\begin{lemma}[Exceptional Items Lower Bound] Let $X_1,X_2,\dots,X_t$ be
	non-negative discrete random variables each taking value zero or at
	least $L$. If $\sum_j \E[X_j]\ge L$ then $\E[\max_j X_j]\ge
	L/2$. \label{lem:KRT-max-sum}
\end{lemma}
\begin{pf} The Bernoulli case of this lemma appears as~\cite[Lemma
	3.3]{KRT}. The extension to the general case is easy. For each $X_j$,
	introduce independent {\em Bernoulli} random variables $\{X_{jk}\}$
	where each $X_{jk}$ corresponds to a particular instantiation $s_{jk}$
	of $X_j$, i.e. $\Pr[X_{jk} = s_{jk}] = \Pr[X_{j } = s_{jk}]$. Note
	that $\max_k X_{jk}$ is stochastically dominated by $X_j$: so
	$\E[\max_j X_j] \ge \E[\max_{jk} X_{jk}]$. Moreover, $\sum_{jk}
	\E[X_{jk}] = \sum_j \E[X_j]\ge L$. So the lemma follows from the
	Bernoulli case.
\end{pf}

\subsection{Effective Size and Its Properties}
\label{sec:effective-size}

As is often the case for stochastic optimization problems, we want to find some
deterministic quantity that is a good surrogate for each random
variable, and then use this deterministic surrogate instead of the
actual random variable. 
Here, we use  the \emph{effective size}, which is
based on the logarithm of the (exponential) moment generating
function~\cite{Hui,Kelly-notes,ElwaM}. 
\begin{definition}[Effective Size]
	\label{defn:logmgf}
	For any random variable $X$ and integer $k\ge 2$, define
	\begin{gather}
		\beta_k(X) \,\, :=\,\, \frac{1}{\log k} \cdot \log \E\left[
		e^{(\log k) \cdot X}\right].
	\end{gather}
	Also define $\beta_1(X) := \E[X]$.
\end{definition}
To get some intuition for this, consider 
independent r.v.s $Y_1, \ldots, Y_n$.
Then if $\sum_i \beta_k(Y_i) \leq b$, 
\begin{align*}
\Pr[ \sum_i Y_i \geq c ] &= \Pr[ e^{\log k \sum_i Y_i} \geq
e^{(\log k ) c} ]\leq \frac{\E[ e^{\log k \sum_i Y_i} ]}{e^{(\log k) c}}  = \frac{\prod_i \E[ e^{(\log k) Y_i} ]}{e^{(\log
		k) c}}
\end{align*}
Taking logarithms, we get
\begin{align*}
&\log \Pr[ \sum_i Y_i \geq c ] \leq \log k \cdot \bigg[ \sum_i
\beta_k(Y_i) - c \bigg]\ \implies \ \Pr[ \sum_i Y_i \geq c ]
\leq \frac1{k^{c-b}}.
\end{align*}

The above calculation, very reminiscent of the standard Chernoff bound
argument, can be summarized by the following lemma (shown, e.g.,
in~\cite{Hui}).
\begin{lemma}[Upper Bound]
	\label{lem:chern}
	For independent random variables $Y_1, \ldots, Y_n$, if $\sum_i \beta_k(Y_i) \leq 	b$ then $\Pr[ \sum_i Y_i \geq c ] \leq (1/k)^{c-b}$.
\end{lemma}
The usefulness of this definition comes from a partial converse, proved in \cite{KRT}:
\begin{lemma}[Lower Bound]
	\label{lem:conv}
	Consider  independent  Bernoulli random variables $Y_1, \ldots, Y_n$ where each $Y_i$ has
	non-zero size $s_i$ being an inverse power of $2$ such that $1/(\log
	k) \leq s_i \leq 1$. If $\sum_i \beta_k(Y_i) \geq 7$ then $\Pr[
	\sum_i Y_i \geq 1 ] \geq 1/k$.
\end{lemma}

\paragraph{Outline of the algorithm for identical machines}
In using the effective size, it is important to set the parameter $k$
carefully.  For  identical machines \cite{KRT}   used $k = m$ the total number of machines. Using the facts discussed above, we can now  outline  their  algorithm/analysis (assuming that all r.v.s are truncated). If the total effective size is at most (say) $20m$ then the jobs can be assigned to $m$ machines in a way that the effective-size load on each machine is at most $21$. By Lemma~\ref{lem:chern} and union bound, it follows that the probability
of some machine exceeding load $23$ is at most $m \cdot (1/m)^{2}=1/m$. On the other hand,  if the total effective size is more than $20m$ then even if the solution was to  balance these evenly, each machine would have effective-size load  at least $7$. By Lemma~\ref{lem:conv} it follows that the load on each machine exceeds one   with
probability $1/m$, and so with $m$ machines this gives a certificate
that the makespan is $\Omega(1)$.

\paragraph{Challenges with unrelated machines}
For unrelated machines, this kind of argument breaks down even in the restricted-assignment setting  where each job can go on only some subset of machines. This is because we don't know what probability of success we want to aim for.
For example, even if the machines had the same speed, but there were
jobs that could go only on $\sqrt{m}$ of these machines, and others
could go on the remaining $m - \sqrt{m}$ of them, we would want their
effective sizes to be quite different. (See the example below.)  And once we go to general unrelated
machines, it is not clear if  any combinatorial argument would suffice. Instead, we propose an LP-based lower bound that  enforces one such constraint (involving effective sizes) for every subset of machines.

\paragraph{Bad Example for Simpler Effective Sizes}
For stochastic load balancing on identical machines~\cite{KRT} showed that any algorithm which maps
each r.v. to a single real value and performs load balancing on these (deterministic) values incurs an
$\Omega(\frac{\log m}{\log\log m})$ approximation ratio. This is precisely the reason they introduced the notion of truncated and exceptional r.v.s. For truncated r.v.s, their algorithm showed that it suffices to use $\beta_{m}(X_j)$ as the deterministic value 
and perform load balancing with respect to these. Exceptional r.v.s were handled separately (in a simpler manner). For unrelated machines, we now  
provide an example which shows that even when all r.v.s are truncated, any algorithm which maps each r.v. to a single real value must incur approximation ratio at least $\Omega(\frac{\log m}{\log\log m})$. This suggests that more work is needed to define the “right” effective sizes in the unrelated machine setting.

There are $m$ machines and $m + \sqrt{m}$ jobs. Each r.v. $X_j$ takes value 1 with probability $\frac{1}{\sqrt{m}}$ (and 0 otherwise). The first $\sqrt{m}$ jobs  can only be assigned to machine 1. The remaining $m$ jobs can be assigned to any machine. Note that $OPT \approx 1 + 1/e$ which is obtained by assigning the first $\sqrt{m}$ jobs to machine 1, and each of the remaining $m$ jobs in a one-to-one manner.
Given any fixed mapping of r.v.s to reals, note that all the $X_j$ get the same value (say $\theta$) as they are identically distributed. So the optimal value of the corresponding (deterministic) load balancing instance is $\sqrt{m}\cdot\theta$. Hence the solution which maps $\sqrt{m}$ jobs to each of the first $1 + \sqrt{m}$ machines is an optimal solution to the deterministic instance. However,   the expected makespan of this assignment is $\Omega(\frac{\log m}{\log\log m})$.

\medskip

We will use the following specific result  in dealing with truncated r.v.s.

\begin{lemma}[Truncated Items Lower Bound] 
	\label{lem:krt-LB} Let $X_1,X_2,\cdots X_n$ be independent $[0,1]$
	r.v.s, and $\{J_i\}_{i=1}^m$ be any partition of
	$[n]$. 
	If $\sum_{j=1}^n \beta_{m}(X_j)\ge 17m$ then $\E\left[\max_{i=1}^m
	\sum_{j\in J_i} X_j\right]=\Omega(1)$.
\end{lemma}
\begin{pf} 
	This is a slight extension of \cite[Lemma~3.4]{KRT}, with two main
	differences.  Firstly, we want to consider arbitrary instead of just
	Bernoulli r.v.s.  Secondly, we use a different definition of effective
	size than they do. We provide the details below for
	completeness. 
	
	At the loss of factor two in the makespan, we may assume (by rounding
	down) that the only values taken by the $X_j$ r.v.s are inverse powers
	of $2$. For each r.v. $X_j$, applying \cite[Lemma~3.10]{KRT} yields
	independent Bernoulli random variables $\{Y_{jk}\}$ so that for each
	power-of-$2$ value $s$ we have
	$$\Pr[X_j = s]=\Pr[s\le \sum_k Y_{jk} <2s].$$ 
	Let $\overline{X}_j=\sum_k Y_{jk}$, so $X_j\le \overline{X}_j< 2\cdot
	X_j$ and $\beta_{m}(\overline{X}_j) = \sum_k
	\beta_{m}(Y_{jk})$. Note also that $\beta_{m}( \overline{X}_j) \ge
	\beta_{m}(X_j)$. Hence $\sum_{jk} \beta_{m}(Y_{jk})\ge
	\sum_{j=1}^n \beta_{m}(X_j) \ge 17m$. Now, consider the assignment
	of the $Y_{jk}$ r.v.s corresponding to $\{J_i\}_{i=1}^m$, i.e. for
	each $i\in [m]$ and $j\in J_i$, all the $\{Y_{jk}\}$ r.v.s are
	assigned to part $i$. Then applying \cite[Lemma~3.4]{KRT} which works
	for Bernoulli r.v.s, we obtain $\E\left[\max_{i=1}^m \sum_{j\in J_i}
	\sum_k Y_{jk} \right]=\Omega(1)$. Observe that the above lemma used
	a different notion of effective size: $\beta'_{1/m}(X) :=
	\min\{s,sqm^s\}$ for any Bernoulli r.v.\ $X$ taking value $s$ with
	probability $q$. However, as shown in \cite[Prop~2.5]{KRT},
	$\beta_{m}(X)\le \beta'_{1/m}(X)$ which implies the version that we
	use here.
	
	Finally, using $X_j>\frac12 \overline{X}_j$ we obtain
\begin{align*}
\E\left[\max_{i=1}^m \sum_{j\in J_i} X_j\right] &\ge \frac12 \E\left[\max_{i=1}^m \sum_{j\in J_i}\overline{X}_j\right]= \frac12 \E\left[\max_{i=1}^m \sum_{j\in J_i} \sum_k Y_{jk} \right]=\Omega(1),
\end{align*}
which completes the proof. 
\end{pf}

\subsection{Non-Adaptive and Adaptive Solutions} We note that our model involves computing an assignment  that is  fixed {\em a priori}, before observing any random instantiations. Such solutions are commonly called non-adaptive. A different class of solutions (called adaptive) involves assigning jobs to machines sequentially, observing the random instantiation of each assigned job.   Designing  approximation algorithms for  the adaptive and non-adaptive models are mutually incomparable. For makespan minimization on identical machines, Graham’s list scheduling already gives a trivial 2-approximation algorithm in the \emph{adaptive} case (in fact, it is 2-approximate on an per-instance basis), whereas the \emph{non-adaptive} case is quite non-trivial and  the Kleinberg et al.~\cite{KRT} result was the first constant-factor approximation. 


We now provide an instance with  identical machines where there is an
$\Omega({\frac{\log m}{\log\log m}})$ gap between the best non-adaptive assignment (the setting of this paper) and the best adaptive assignment. 
The instance consists of $m$ machines and $n = m^2$ jobs each of which is identically distributed taking size 1 with probability $\frac{1}{m}$ (and 0 otherwise). Recall that Graham's algorithm considers jobs in any order and places each job on the least loaded machine. It follows that the expected makespan of this adaptive policy is at most $1 + \frac{1}{m}\cdot\E[X_j]=2$. On the other hand, the best static assignment has expected makespan $\Omega({\frac{\log m}{\log\log m}})$, which is obtained by assigning $m$ jobs to each machine.

\subsection{Useful Probabilistic Inequalities}
\begin{theorem}[Jensen's Inequality] Let $X_1,X_2,\dots,X_t$ be  random 
	variables and $f(x_1,\cdots, x_t)$ any convex function. Then $$\E[f(X_1,\cdots ,X_t)] \ge f(\E[X_1],\cdots ,\E[X_t]).$$ \label{thm:jensen}
\end{theorem}

\begin{theorem}[Rosenthal Inequality]\emph{~\cite{Rosen,JSZ,Latala}}
	\label{thm:rosen}  
	Let $X_1,X_2,\dots,X_t$ be  independent non-negative random
	variables. Let $q\geq1$ and $K =\Theta(q/\log q)$. Then it is the case that
{\small	\begin{align*}
	\E\bigg[\bigg(\sum_j{X_j}\bigg)^q\bigg] \le K^q \cdot \max\left\{
	\left(\sum_j\E[X_j]\right)^{q},\sum_{j}\E[X_j^q]
	\right \}.
	\end{align*} }
\end{theorem}


\section{Makespan Minimization}
\label{sec:makespan}

The main result of this section is:
\MMThm*

Using a binary search scheme and scaling, it suffices to find one of the
following:
\begin{itemize}
	\item[(i)] \emph{upper bound:} a solution with expected makespan at most
	$O(1)$, or
	\item[(ii)] \emph{lower bound:} a certificate that the optimal expected
	makespan is more than one.
\end{itemize}
Hence, we assume that the optimal solution for the instance has unit
expected makespan, and try to find a solution with expected makespan $b
= O(1)$; if we fail we output a lower bound certificate.

At a high level, the ideas we use are the following: first, in
\S\ref{sec:new-lower-bound} we show a more involved lower bound based on
the effective sizes of jobs assigned to every subset of machines. This
is captured using an exponentially-sized LP which is solvable in
polynomial time. Then, to show that this lower bound is a good one, we
give a new rounding algorithm for this LP in \S\ref{sec:rounding} to get
an expected makespan within a constant factor of the lower bound.

\subsection{A New Lower Bound}
\label{sec:new-lower-bound}

Our starting point is a more general lower bound on the makespan. The
(contrapositive of the) following lemma says that if the effective sizes
are large then the expected makespan must be large too. This is much the
same spirit as Lemma~\ref{lem:conv}, but for the general setting of
unrelated machines.


\begin{lemma}[New Valid Inequalities]
	\label{lem:stoch-lb} 
	Consider any feasible solution that assigns jobs $J_i$ to each machine
	$i\in [m]$. If the expected makespan $\E\left[\max_{i=1}^m \sum_{j\in
		J_i} X_{ij}\right] \le 1$, then
	\begin{gather}
		\sum_{i=1}^m\sum_{j\in J_i} \E[X''_{ij}] \le 2, \qquad \mbox{ and
		} \label{eq:new-lb-1} \\
		\sum_{i\in K}  \, \sum_{j\in J_i}
		\, \beta_{k}(X'_{ij}) \,\, \le \,\, O(1)\cdot k,\ \mbox{ for all }K\sse [m], \quad \mbox{where
		}k=|K|. \label{eq:new-lb-2}
	\end{gather}
\end{lemma}
\begin{pf}
	The first inequality~\eqref{eq:new-lb-1} focuses on the exceptional
	parts, and loosely follows from the intuition that if the sum of
	biases of a set of independent coin flips is large (exceeds $2$ in
	this case) then you expect one of them to come up heads. Formally, the
	proof follows from Lemma~\ref{lem:KRT-max-sum} applied to
	$\{X''_{ij}\, :\, j\in J_i, i\in[m]\}$.  
	
	For the second inequality~\eqref{eq:new-lb-2}, consider any subset
	$K\sse[m]$ of the machines. Then the total effective size of the jobs
	assigned to these machines must be small, where now the effective size $\beta_k$ 
	can be measured with parameter $k =  |K|$. Formally applying
	Lemma~\ref{lem:krt-LB} only to the $k$ machines in $K$ and the \normal
	random variables $\{X'_{ij} : i\in K, j\in J_i\}$ corresponding to
	jobs assigned to these machines, we obtain the desired inequality.
\end{pf}

Given these valid inequalities, our algorithm now seeks an assignment
satisfying~\eqref{eq:new-lb-1}--\eqref{eq:new-lb-2}. If we fail, the
lemma assures us that the expected makespan must be large. On the other
hand, if we succeed, such a ``good'' assignment by itself is not
sufficient.  The challenge is to show the converse of
Lemma~\ref{lem:stoch-lb}, i.e., that any assignment
satisfying~\eqref{eq:new-lb-1}--\eqref{eq:new-lb-2} gives us an expected
makespan of $O(1)$.
 
Indeed, towards this goal, we first write an LP relaxation with an
exponential number of constraints, corresponding
to~\eqref{eq:new-lb-2}. We can solve this LP using the ellipsoid
method. Then, instead of rounding the fractional solution to satisfy all
constraints (which seems very hard), we show how to satisfy only a
carefully chosen subset of the constraints~\eqref{eq:new-lb-2} so that
the expected makespan can still be bounded. Let us first give the LP
relaxation.

In the ILP formulation of the above lower bound, we have binary
variables $y_{ij}$ to denote the assignment of job $j$ to machine $i$,
and fractional variables $z_i(k)$ denote the total load on machine $i$
in terms of the deterministic effective sizes
$\beta_{k}$. Lemma~\ref{lem:stoch-lb} shows that the following
feasibility LP is a valid relaxation:
\begin{align}
	\sum_{i=1}^m y_{ij} &= 1,\,\, \forall j\in
	[n], \label{eq:LP:assign}\\ 
	z_i(k) - \sum_{j=1}^n  \beta_{k}(X'_{ij}) \cdot y_{ij} &= 0,\,\,\forall i\in [m], \  \forall k=1,2,\cdots m, \label{eq:LP:load} \\ 
	\sum_{i=1}^{m} \sum_{j=1}^n  \E[X''_{ij}]  \cdot y_{ij}  &\le
	2, \label{eq:LP:exceptn} \\ 
	\sum_{i\in K} z_i(k) &\le b\cdot k,\,\,\forall K\sse [m] \mbox{ with }|K|=k,\ \forall k=1,2,\cdots m,\label{eq:LP:subset} \\ 
	y_{ij}, z_i(k)  &\ge 0,\quad\forall i,j,k. \label{eq:LP:nonneg}
	\end{align}
In the above LP, $b=O(1)$ denotes the constant multiplying $k$ in the right-hand-side
of~\eqref{eq:new-lb-2}.

Although this LP has an exponential number of constraints (because
of~\eqref{eq:LP:subset}), we can give an efficient separation
oracle. Indeed, consider a candidate solution $(y_{ij}, z_i(k))$, and
some integer $k$; suppose we want to verify~\eqref{eq:LP:subset} for
sets $K$ with $|K|=k$. We just need to look at the $k$ machines with the
highest $z_i(k)$ values and check that the sum of $z_i(k)$ for these
machines is at most $bk$. So, using the Ellipsoid method we can assume
that we have an optimal solution $(y,z)$ for this LP in polynomial time.
We can summarize this in the following proposition:

\begin{proposition}[Lower Bound via LP]
	\label{fct:lb}
	The linear program~(\ref{eq:LP:assign})--(\ref{eq:LP:nonneg}) can be
	solved in polynomial time. Moreover, if it is infeasible, then the
	optimal expected makespan is more than $1$.
\end{proposition}

\subsection{The Rounding}
\label{sec:rounding}

\paragraph{Intuition.} In order to get some intuition about the rounding
algorithm, let us first consider the case when the assignment variables
$y_{ij}$ are either 0 or 1, i.e., the LP solution assigns each job
integrally to a machine. In order to bound the expected makespan of this solution, let $Z_j$ denote the variable $X_{ij}$, where
$j$ is assigned to $i$ by this solution. First consider the exceptional parts  $Z''_j$ of the 
random variables. Constraint~\eqref{eq:LP:exceptn} implies that $\sum_{j }
\E[Z''_j]$ is at most 2. Even if the solution assigns all of these jobs to
the same machine, the contribution of these jobs to the expected
makespan is at most $\sum_{j } \E[Z''_j]$, and hence at most 2. Thus,
we need only worry about the \normal  $Z'_j$ variables. 

Now for a machine $i$ and integer $k \in [m]$, let $z_i(k)$ denote the
sum of the effective sizes $\beta_{k}(Z'_j)$ for the \normal  r.v.s 
assigned to $i$. We can use Lemma~\ref{lem:chern} to infer that if
$z_i(m) = \sum_{j \text{ assigned to } i} \beta_{m}(Z_j) \leq b$, then
the probability that these jobs have total size at most $b+2$ is at
least $1 - 1/m^2$.  Therefore, if $z_i(m) \leq b$ for all machines $i
\in [m]$, then by a trivial union bound the probability that makespan is
more than $b+2$ is at most $1/m$. Unfortunately, we are not done. All we
know from constraint~\eqref{eq:LP:subset} is that the \emph{average}
value of $z_i(m)$ is at most $b$ (the average being taken over the $m$
machines). However, there is a clean solution. It follows that there is
at least one machine $i$ for which $z_i(m)$ is at most $b$, and so the
expected load on such machines stays $O(1)$ with high probability. Now
we can ignore such machines, and look at the residual problem. We are
left with $k < m$ machines. We recurse on this sub-problem (and use the
constraint~\eqref{eq:LP:subset} for the remaining set of machines). The
overall probability that the load exceeds $O(1)$ on any machine can then
be bounded by applying a union bound.

Next,   we  address the fact that $y_{ij}$ may be not be
integral. It seems very difficult to round a fractional solution while respecting all the (exponentially many) constraints in~\eqref{eq:LP:subset}. Instead, we observe that the  expected makespan analysis (outlined above) only utilizes a linear number of constraints in~\eqref{eq:LP:subset}, although this subset is not known {\em a priori}. Moreover,  for each machine $i$, the above analysis  only uses $z_i(k)$  for a {\em single}  
value of $k$ (say $k_i$).    Therefore, it suffices to find an 2 integral  assignment  that bounds the load of each machine $i$ in terms of effective sizes $\beta_{k_i}$. 
It turns out that this problem is precisely  an instance of the Generalized Assignment Problem (\GAP), for which we utilize the algorithm from~\cite{ShmoysT93}. 

\paragraph{The Rounding Procedure.} We now describe the iterative
procedure formally. Assume we have an LP solution $\{y_{ij}\}_{i \in
	[m], j \in [n]}, \{ z_i(k)\}_{i,k \in [m]}$.
\begin{enumerate}
	\item Initialize $\ell\gets m$, $L\gets [m]$,
	$c_{ij}\gets\E[X''_{ij}]$.
	\item While $(\ell > 0)$ do:
	\begin{enumerate}
		\item Set $L'\gets \{i\in L : z_i(\ell) \le b\}$. Machines in $L'$ are said to be in   {\em class $\ell$}. 
		\item Set $p_{ij}\gets \beta_{\ell}(X'_{ij})$ for all $i\in L'$ and $j\in [n]$.
		\item Set $L\gets L\setminus L'$ and $\ell=|L|$.
	\end{enumerate}
	\item Define a deterministic instance \I of the \GAP as follows: the set
	of jobs and machines remains unchanged. For each job $j$ and machine
	$i$, define $p_{ij}$ and $c_{ij}$ as above. The makespan bound is
	$b$. Use the algorithm of Shmoys and Tardos~\cite{ShmoysT93} to find
	an assignment of jobs to machines. Output this solution.
\end{enumerate}

Recall that in an
instance $\I$ of \GAP, we are given a set of $m$ machines and $n$
jobs. For each job $j$ and machine $i$, we are given two quantities:
$p_{ij}$ is the processing time of $j$ on machine $i$, and $c_{ij}$ is
the cost of assigning $j$ to $i$. We are also given a makespan bound
$b$. Our goal is to assign jobs to machines to minimize the total cost
of assignment, subject to the total processing time of jobs assigned to each
machine being at most $b$. If the  natural   LP relaxation
for this problem has optimal value  $C^\star$ then the
algorithm in~\cite{ShmoysT93}  finds in 
polynomial-time an assignment with cost  
at most $C^\star$ and  makespan is at most $B+\max_{i,j} p_{ij}$.

\subsection{The Analysis}

We begin with some simple observations:
\begin{observation}\label{obs:terminate}
	The above rounding procedure terminates in at most $m$ iterations.
	Furthermore, for any $1\le \ell \le m$, there are at most $\ell$
	machines of class at most $\ell$.
\end{observation}
\begin{pf}
	The first statement follows from the fact that $L'\ne \emptyset$ in
	each iteration. To see this, consider any iteration involving a set
	$L$ of $\ell$ machines. The LP constraint~\eqref{eq:LP:subset} for $L$
	implies that $\sum_{i\in L} z_i(k) \le b\cdot \ell$, which means there
	is some $i\in L$ with $z_i(\ell)\le b$, i.e., $L'\ne \emptyset$.  The
	second statement follows from the rounding procedure: the machine
	classes only decrease over the run of the algorithm, and the class assigned to any unclassified machine equals the current 
	number of  unclassified machines.
\end{pf}

\begin{observation}\label{obs:det-lp}
	The solution $y$ is a feasible fractional solution to the natural LP
	relaxation for the \GAP instance \I. This solution has makespan at
	most $b$ and fractional cost at most 2. The rounding algorithm of
	Shmoys and Tardos~\cite{ShmoysT93} yields an assignment with makespan
	at most $b+1$ and cost at most 2 for the instance $\I$.
\end{observation}
\begin{pf}
	Recall that the natural LP relaxation is the following:
	\begin{alignat}{2}
		\min~~~ \textstyle \sum_{ij} c_{ij} y_{ij} &&& \notag\\
		\textstyle \sum_{j} p_{ij} y_{ij} & \leq b, & \qquad\qquad& \forall
		i, \label{eq:makesp} \\
		\textstyle \sum_i y_{ij} &= 1, & & \forall j, \label{eq:assign} \\
		y_{ij} &= 0, & & \forall j~ \mbox{s.t.}~ p_{ij} > 1, \label{eq:verboten} \\
		y & \geq 0. & & \notag
	\end{alignat}
	Firstly, note that by~\eqref{eq:LP:assign}, $y$ is a valid fractional
	assignment that assigns each job to one machine, which
	satisfies~(\ref{eq:assign}). 
	
	Next we show~(\ref{eq:makesp}), i.e., that $\max_{i=1}^m \sum_{j=1}^n
	p_{ij}\cdot y_{ij} \le b$. This follows from the definition of the
	deterministic processing times $p_{ij}$. Indeed, consider any machine
	$i\in [m]$. Let $\ell$ be the class of machine $i$, and $L$ be the
	subset of machines in the iteration when $i$ is assigned class $\ell$. This means that
	$p_{ij}=\beta_{\ell}(X'_{ij})$ for all $j\in [n]$. Also, because
	machine $i\in L'$, we have $z_i(\ell) = \sum_{j=1}^n
	\beta_{\ell}(X'_{ij}) \cdot y_{ij} \le b$. So we have $\sum_{j=1}^n
	p_{ij}\cdot y_{ij} \le b$ for each machine $i\in [m]$. 
	
	Finally, since the random variable $X'_{ij}$ is at most 1, we get that
	for any parameter $k\ge 1$, $\beta_k(X'_{ij}) \leq 1$; this implies
	that $p_{max} := \max_{i,j} p_{ij} \leq 1$ and hence the constraints~(\ref{eq:verboten}) are
	vacuously true. Finally, by~\eqref{eq:LP:exceptn}, the objective
	is $\sum_{i=1}^{m} \sum_{j=1}^n c_{ij} \cdot
	y_{ij}=\sum_{i=1}^{m} \sum_{j=1}^n \E[X''_{ij}] \cdot y_{ij} \le 2$.
	Therefore the rounding algorithm~\cite{ShmoysT93}
	yields an assignment of makespan at most $b+p_{\max} \leq b+1$, and
	of cost at most 2.
\end{pf}

In other words, if $J_i$ be the set of jobs assigned to machine $i$ by
our algorithm, Observation~\ref{obs:det-lp} shows that this assignment
has the following properties (let $\ell_i$ denote the class of machine
$i$):

\begin{align}
	\sum_{i=1}^m\sum_{j\in J_i} \E[X_{ij}''] \,\, &\le \,\, 2, \quad and 
	\label{eq:stoch-cost-bound}
	\\
	\sum_{j\in J_i} \beta_{\ell_i}(X_{ij}') \,\, &\le \,\, b+1,\qquad
	\forall i\in[m]. \label{eq:stoch-mgf-bound} 
\end{align}
Note that we ideally wanted to give an assignment that satisfied
(\ref{eq:new-lb-1})--(\ref{eq:new-lb-2}), but instead of giving a bound
for all sets of machines, we give just the bound on the
$\beta_{\ell_i}$ values of the jobs for each machine $i$. The next
lemma shows this is enough.

\begin{lemma}[Bounding the Makespan]
	\label{lem:final}
	The expected makespan of the assignment $\{J_i\}_{i \in [m]}$ is at
	most $4b+10$.
\end{lemma}

\newcommand{\Ihigh}{I^{\text{hi}}}

\begin{pf}
	Let $\Ihigh$ denote the index set of machines of class~3 or
	higher. Observation~\ref{obs:terminate} shows that there are at most 3
	machines which are not in $\Ihigh$. For a machine $i$, let $T_i = \sum_{j
		\in J_i} X_{ij}'$ denote the
	total load due to \normal sizes of jobs assigned to it. Clearly, the makespan is bounded by
	$$ \max_{i \in \Ihigh} T_i  + 
	\sum_{i \notin \Ihigh} T_i + \sum_{i=1}^m \sum_{j \in J_i} X_{ij}''. $$ 
	The
	expectation of third term is at most two,
	using~\eqref{eq:stoch-cost-bound}. We
	now bound the expectation of the second  term above.  A direct application of Jensen's
	inequality (Theorem~\ref{thm:jensen}) for concave functions shows that $\beta_k(X) \geq \E[X]$
	for any random variable $X$ and any $k \ge 1$. Then applying
	inequality~\eqref{eq:stoch-mgf-bound} shows that $\E[T_i] \leq b+1$
	for any machine $i$. Therefore, the expected makespan of our solution
	is at most
	\begin{equation}
		\label{eq:makespanfinal}
		\E\left[\max_{i \in \Ihigh} T_i\right] + 3(b + 1) + 2. 
	\end{equation}
	It remains to bound the first term above.
	
	\begin{observation}\label{obs:machine-tail}
		For any machine $i$, $\Pr\left[ \sum_{j\in J_i} X'_{ij} > b+1+\alpha
		\right] \le \ell_i^{-\alpha}$ for all $\alpha\ge 0$.
	\end{observation}
	\begin{pf}
		Inequality~\eqref{eq:stoch-mgf-bound} for machine $i$ shows that
		$\sum_{j\in J_i} \beta_{\ell_i}(X'_{ij}) \le b+1$. But recalling
		the definition of the effective size (Definition~\ref{defn:logmgf}), the result
		follows from Lemma~\ref{lem:chern}.
	\end{pf}
	
	Now we can bound the probability of any machine in $ \Ihigh$ having a
	high makespan.
	\begin{lemma} For any $\alpha>2$, $$\Pr\left[ \max_{i \in \Ihigh} T_i >
		b+1+\alpha \right] \le 2^{2-\alpha}/(\alpha-2).$$
	\end{lemma}
	
	\begin{pf}
		Using a union bound, we get
		\begin{align*}
			\Pr\left[ \max_{i \in \Ihigh} T_i > b+1+\alpha \right] & \le
			\sum_{\ell=3}^m \,\, \sum_{i: \ell_i=\ell} \Pr\left[ T_i >
			b+1+\alpha  \right] \\
			&\le \sum_{\ell=3}^m \ell^{-\alpha}\cdot (\mbox{\# of class
				$\ell$ machines})\\
			&\le \sum_{\ell=3}^m \ell^{-\alpha+1} \,\,   \le\,\,   \int_{x=2}^\infty x^{-\alpha+1} dx \,\,  =\, \frac{2^{-\alpha+2}}{\alpha-2}.
		\end{align*}
		The first inequality uses a trivial union bound, the second uses
		Observation~\ref{obs:machine-tail} above, and the third inequality
		is by Observation~\ref{obs:terminate}.
	\end{pf}
	
	Using the above lemma, we get
	\begin{align*}
		\E[\max_{i \in \Ihigh} T_i] &= (b+4)+\int_{\alpha=3}^\infty
		\Pr[\max_{i \in \Ihigh} T_i > b+1+\alpha] d\alpha\,\, \le\,\, (b+4)+\int_{\alpha=3}^\infty 2^{2-\alpha}d\alpha \,\, \le \,\, b+5.
	\end{align*}
	Inequality~\eqref{eq:makespanfinal} now shows that the expected
	makespan is at most $(b+5) + 3(b+1) + 2$.
\end{pf}

This completes the proof of Theorem~\ref{thm:main}. 

\section{Budgeted Makespan Minimization}
\label{sec:budget-short}

We now consider a generalization of the \stocload problem, called
\budgstoc, where each job $j$ also has \emph{reward} $\reward_j \geq
0$. We are required to schedule some subset of jobs whose total reward is
at least some target value $R$. The objective, again, is to minimize the
expected makespan. If the target $R = \sum_{j \in [n]} \reward_j$ then
we recover the \stocload problem. We 
show: 
\BMMThm*

Naturally, our algorithm/analysis will build on the ideas developed in
\S\ref{sec:makespan}, but we will need some new ideas to handle the fact
that only a subset of jobs need to be scheduled. As in the case of
\stocload problem, we can formulate a suitable LP relaxation. A similar
rounding procedure reduces the stochastic problem to a deterministic
problem, which we call  \budgGAP. An instance of \budgGAP is
similar to that of \GAP, besides the additional requirement that jobs
have rewards and we are required to assign jobs of total reward at least
some target $R$. Rounding the natural LP relaxation for \budgGAP turns
out to be non-trivial. Indeed, using ideas from~\cite{ShmoysT93}, we
reduce this rounding problem to rounding a fractional matching solution
with additional constraints, and solve the latter using polyhedral
properties of bipartite matching polyhedra. 

As before, using a
binary-search scheme (and by scaling down the sizes), we can assume that
we need to either (i)~find a solution of expected makespan $O(1)$, or
(ii)~prove that the optimal value is more than $1$. We use a natural
LP relaxation which has variables $y_{ij}$ for each job $j$ and machine
$i$. The LP includes the
constraints~\eqref{eq:LP:load}-\eqref{eq:LP:nonneg} for the base
problem, and in addition it has the following two constraints:
\begin{alignat}{2}
  \sum_{i=1}^m y_{ij} &\le 1,& \qquad\qquad &\forall j=1,\cdots
  n, \label{eq:atmostone} \\
  \sum_{j=1}^n \reward_j\cdot \sum_{i=1}^m y_{ij}&\ge R. && \label{eq:reward}
\end{alignat}
The first constraint~(\ref{eq:atmostone}) replaces
constraint~\eqref{eq:LP:assign} and says that not all jobs need to be
assigned. The second constraint~(\ref{eq:reward}) ensures that the
assigned jobs have total reward at least the target $R$. For technical reasons that will be clear later, we also perform
a preprocessing step: for $i,j$ pairs where
$\E[X''_{ij}] > 2$, we force the associated $y_{ij}$ variable to zero. Note that by Lemma~\ref{lem:stoch-lb}, this variable fixing is valid for any integral assignment that has expected makespan at most one (in fact, we have  $\sum_i \sum_{j} \E[X''_{ij}]\cdot y_{ij} \leq 2$ for such an assignment).
  As in
\S\ref{sec:new-lower-bound} this LP can be solved in polynomial time via
the ellipsoid method. If the LP is infeasible we get a proof (using
Lemma~\ref{lem:stoch-lb}) that the optimal expected makespan is more
than one. Hence we may assume the LP is feasible, and proceed to round
the solution along the lines of~\S\ref{sec:rounding}.

Recall that the rounding algorithm in~\S\ref{sec:rounding} reduces the
fractional LP solution to an instance of the generalized assignment
problem (\GAP). Here, we will use a further generalization of \GAP,
which we call \budgGAP. An instance of this problem is similar to an
instance of \GAP. We are given $m$ machines and $n$ jobs, and for each
job $j$ and machine $i$, we are given the processing time $p_{ij}$ and
the associated assignment cost $c_{ij}$. Now each job $j$ has a reward
$\reward_j$, and there are two ``target'' parameters: the reward target
$R$ and the makespan target $B$. We let $p_{max}$ and $c_{max}$ denote the maximum values of  processing time and cost respectively. 
A solution must assign a subset of jobs
to machines such that the total reward of assigned jobs is at least $R$.
Moreover, as in the case of \GAP, the goal is to minimize the total
assignment cost subject to the condition that the makespan is at most
$B$.  Our main technical theorem of this section shows how to round an
LP relaxation of this \budgGAP problem.
\begin{theorem}\label{thm:bGAP}
  There is a polynomial-time rounding algorithm for \budgGAP that given
  any fractional solution to the natural LP relaxation of cost $C^*$,
  produces an integer solution having total cost at most $C^*+c_{max}$
  and makespan at most $B+2p_{max}$.
\end{theorem}

Before we prove this theorem, let us use it to solve the \budgstoc, and
prove Theorem~\ref{thm:main-reward}. Proceeding as in~\S\ref{sec:rounding}, we
perform Steps~1-2 from the rounding procedure. This rounding gives us
values $p_{ij}$ and $c_{ij}$ for each job/machine pair. Now, instead
of reducing to an instance of \GAP, we reduce to an instance $\I'$ of
\budgGAP.  The instance $\I'$ has the same set of jobs and machines as
in the original \budgstoc instance $\I$.  For each job $j$ and machine
$i$, the processing time and the assignment cost are given by $p_{ij}$
and $c_{ij}$ respectively.  Furthermore, the reward $\reward_j$ for job
$j$, and the reward target $R$ are same as those in $\I$.  The makespan
bound $b = O(1)$ (as in~\eqref{eq:LP:subset}). It is easy to check that
the fractional solution $y_{ij}$ is a feasible fractional solution to
the natural LP relaxation for $\I'$ (given below), and the assignment
cost of this fractional solution is at most 2. Applying
Theorem~\ref{thm:bGAP} yields an assignment $\{J_i\}_{i=1}^m$, which has
the following properties:
\begin{itemize}
\item The makespan is at most $b+2= O(1)$; i.e., $\sum_{j\in J_i} p_{ij}
  \le b + 2p_{max} \leq b+2$ for each machine $i$. Here we used the fact
  that $p_{max}\le 1$.
\item The cost of the solution, $\sum_{i=1}^m \sum_{j\in J_i} c_{ij}$,
  is at most 4. This uses the fact that the LP cost  $C^*=\sum c_{ij}\cdot y_{ij} \le 2$ and
  $c_{max}\le 2$ by the preprocessing on the $\E[X''_{ij}]$ values.
\item The total reward for the assigned jobs, $\sum_{j\in \cup_i J_i}
  \reward_j$, is at least $R$.
\end{itemize}
Now arguing exactly as in \S\ref{sec:rounding},  the first two
properties imply that the expected makespan is $O(1)$.  The third
property implies the total reward of assigned jobs is at least $R$, and
completes the proof of Theorem~\ref{thm:main-reward}. 

\subsection{Proof of Theorem~\ref{thm:bGAP}}

\hfill
\begin{pf}[Proof of Theorem~\ref{thm:bGAP}]
  Let $\I$ be an instance of \budgGAP as described above. The natural LP
  relaxation for this problem is as follows:
  \begin{alignat}{2}
    \min~~~ \textstyle \sum_{ij} c_{ij} y_{ij} &&& \notag\\
    \textstyle \sum_{j} p_{ij} y_{ij} & \leq B, & \qquad\qquad& \forall
    i, \label{eq:makesp1} \\
    \textstyle \sum_i y_{ij} &\leq 1, & & \forall j, \label{eq:assign1} \\
    \textstyle \sum_{i,j} y_{ij}\reward_j &\geq R, & &  \label{eq:wt1} \\
    y_{ij} &= 0, & & \forall j~ \mbox{s.t.}~ p_{ij} > b, \label{eq:verboten1} \\
    y & \geq 0. & & \notag
  \end{alignat}
  
  Let $\{y_{ij}\}$ denote an optimal fractional solution to this LP.
  For each machine $i$, let $t_i := \lceil \sum_j y_{ij} \rceil$ be the
  (rounded) fractional assignment to machine $i$. Using the algorithm in Theorem 2.1 of~\cite{ShmoysT93}, we obtain  a bipartite graph $G = (V_1
  \cup V_2, E)$ and a fractional matching $y'$ in $G$, where:
  \begin{itemize}
      \item $V_1 = [n]$ is the set of jobs and $V_2$ (indexed by $i'=1, \ldots, m'$) consists of $t_i$ copies for each machine $i\in[m]$. The cost $c_{i'j}=c_{ij}$ for any job $j\in [n]$ and any machine-copy $i'$ of machine $i\in[m]$.  
 \item for each job $j\in [n]$ we have  $\sum_{i=1}^{m'} y'_{ij} = \sum_{i=1}^{m} y_{ij} \le 1$ for all $j\in [n]$.
\item the  reward  $\sum_{j=1}^{n} \reward_j \sum_{i=1}^{m'} y'_{ij} \ge R$  and the  
    cost $\sum_{i'=1}^{m'} \sum_{j=1}^n c_{i'j} y'_{ij} = C^*$ are same as for $y$.
  \item   the jobs of $V_1$ incident to copies of any machine $i\in [m]$ can be divided into (possibly overlapping) groups $H_{i,1},\cdots H_{i,t_i}$ where 
  $$\sum_{j\in H_{i,g}} y'_{ij} =1  \,  \mbox{ for all }1\le g\le t_i-1\quad \mbox{ and} \quad  \sum_{j\in H_{i,t_i}} y'_{ij} \le 1,$$
  and for any two consecutive groups $H_{i,g}$ and $H_{i,g+1}$ we have $p_{ij} \ge p_{ij'}$ for all $j\in H_{i,g}$ and $j'\in H_{i,g+1}$.
  Informally, this is achieved by sorting the
  jobs in non-increasing order of $p_{ij}$, and  assigning the $k^{th}$ unit of
  $\sum_j y_{ij}$ to the $k^{th}$ machine-copy for each $1\le k \le t_i$. 
    \end{itemize}
A crucial property of this construction shown in \cite{ShmoysT93} is that any  assignment that places at most one job on each
    machine-copy has makespan at most $B+p_{max}$ in the original
    instance $\I$ (where for every machine $i$, we assign to it all the
    jobs which are assigned to a copy of $i$ in this integral
    assignment).   We will use the following  simple  extension of this  property: if the
   assignment places {\em two} jobs on one machine-copy and at
  most one job on all other machine-copies, then it has makespan at most
  $B+2p_{max}$ in the instance $\I$. 
  
Observe that the  
  solution $y'$ is a feasible solution to the following LP with variables $\{z_{ij}\}_{(i,j) \in E}$. 
  \begin{alignat}{2}
    \min~~~ \textstyle \sum_{ij} c_{ij} z_{ij} &&& \label{eq:bmk-obj}\\
    \textstyle \sum_{i \in [m'] : (ij)\in E} z_{ij} &\leq 1, & \qquad\qquad& \forall j \in [n], \label{eq:assign-aux} \\
    \textstyle \sum_{j \in [n] : (ij)\in E} z_{ij} &\leq 1, & & \forall i \in [m'], \label{eq:assign2-aux} \\
    \textstyle \sum_{(ij)\in E} \reward_j \cdot z_{ij} &\geq R, & &  \label{eq:wt-aux} \\
    z & \geq 0. & & \label{eq:bmk-var}
  \end{alignat}
  So the optimal value of this auxiliary LP is at most $C^*$. We note
  that its integrality gap is unbounded even when $c_{max}$ is small; see the example below. So this differs from \cite {ShmoysT93} for the usual \GAP where the corresponding LP (without \eqref{eq:wt-aux}) is actually integral.   However, we
  show below how to obtain a good integral solution that violates the matching
  constraint for just a {\em single} machine-copy in $V_2$.

  Indeed, let
  $z$ be an optimal solution to this LP: so $c^T z\le C^*$.  Note that the feasible region
  of this LP is just the bipartite-matching polytope on $G$ intersected
  with one extra linear constraint~(\ref{eq:wt-aux}) that corresponds to
  the total reward being at least $R$. So $z$ must be a convex
  combination of two adjacent extreme points of the bipartite-matching
  polytope. Using the integrality and adjacency properties
  (see~\cite{BR74}) of the bipartite-matching polytope, it follows that
  $z=\lambda_1\cdot \mathbf{1}_{M_1} + \lambda_2 \cdot \mathbf{1}_{M_2}$
  where:
  \begin{itemize}
  \item $\lambda_1+\lambda_2=1$ and $\lambda_1,\lambda_2\ge 0$.
  \item $M_1$ and $M_2$ are integral matchings in $G$. 
  \item The symmetric difference $M_1\oplus M_2$ is a single cycle or
    path.
  \end{itemize}
 For any matching $M$ let $c(M)$ and $\reward(M)$ denote its total cost and
  reward respectively.   Without loss of generality, we  assume that $\reward(M_1)\ge \reward(M_2)$. If $c(M_1)\le c(M_2)$ then $M_1$ is itself a solution with reward at least $R$ and cost at most $C^*$. So we assume $c(M_1)>c(M_2)$ below.  
  
  If $M_1\oplus M_2$ is a
  cycle then we output $M_2$ as the solution. Note that the cycle must
  be an even cycle: so the set of jobs assigned by $M_1$ and $M_2$ is
  identical. As the reward function is only dependent on the assigned
  jobs (and not the machines used in the assignment) it follows that
  $\reward(M_2)=\reward(M_1)\ge R$. So $M_2$ is indeed a feasible
  solution and has cost $c(M_2)\le c^T  z \le C^*$.

  Now consider the case that $M_1\oplus M_2$ is a path. If the set of
  jobs assigned by $M_1$ and $M_2$ are the same then $M_2$ is an optimal
  integral solution (as above). The only remaining case is that $M_1$
  assigns one additional job (say $j^*$ to $i^*$) over the jobs in
  $M_2$. Then we return the solution $M_2\cup \{ (j^*,i^*)\}$. Note that this is not  a feasible matching. But the
  only infeasibility is at machine-copy $i^*$ which may have two jobs
  assigned; all other machine-copies have at most one job. The reward of
  this solution is $\reward(M_1)\ge R$. Moreover, its cost is at most
  $c(M_2) + c_{i^*j^*} \le C^* + c_{max}$.  

  Now using this (near-feasible) assignment gives us the desired cost
  and makespan bounds, and completes the proof of
  Theorem~\ref{thm:bGAP}.
\end{pf}
 \medskip\textbf{Integrality Gap for Budgeted Matching LP.} Here we show that the LP~\eqref{eq:bmk-obj}--\eqref{eq:bmk-var} used in the algorithm for budgeted \GAP has an unbounded
integrality gap, even if we assume that $c_{max}\ll OPT$. The instance consists of $n$ jobs and $m = n - 1$
machines. For each machine $i\in [m]$, there are two incident edges in $E$: one to job $i$ (with cost 1) and the other to job $i + 1$ (with cost $n$). So $E$ is the disjoint union of two machine-perfect matchings $M_1$ (of total cost $m$) and $M_2$ (of total cost $mn$). The rewards are
\begin{align*}
    r_j=\begin{cases}1&\mbox{ if }j=1\\
    4&\mbox{ if }2\le j\le n-1\\
    2&\mbox{ if } j = n\end{cases}.
\end{align*}
and the target $R = 4(n-2) + 1 + \epsilon$ where $\epsilon\rightarrow 0$. Note that the only (minimal) integral solution involves assigning the jobs $\{2, 3,\cdots,n\}$ which has total reward $4(n-2) + 2$. This solution has cost $OPT = mn$ and corresponds to matching $M_2$. On the other hand, consider the fractional solution $z=\epsilon \mathbf{1}_{M_2} + (1-\epsilon)\mathbf{1}_{M_1}$. This is clearly feasible for the matching constraints, and its reward is $\epsilon(4(n - 2) + 2) + (1 - \epsilon)(4(n - 2) + 1) = R$. So $z$ is a feasible fractional solution. The cost of this
fractional solution is at most $m + \epsilon(mn)\ll OPT$.

\section{$\ell_q$-norm Objectives}
\label{sec:q-norm}

In this section, we prove Theorem~\ref{thm:main-qnorm}. Given an
assignment $\{J_i\}_{i=1}^m$, the load $L_i$ on machine $i$ is the
r.v.\ 
$L_i := \sum_{j \in J_i}X_{ij}$. Our goal is to find an assignment to
minimize the expected $q$-norm of the load vector $\mathbf{L}:= (L_1,
L_2, \ldots, L_m)$.  Recall that the makespan is $\norm{ \mathbf{L}
}_\infty$ which is  approximated within constant factors by  $\norm{ \mathbf{L}}_{\log m}$. So the $q$-norm problem is a generalization of \stocload.
Our main result here is:
\QNormThm*


We begin by assuming that we know the optimal value~$M$ of the $q$-norm.
Our approach parallels that for the case of minimizing the expected
makespan, with some changes. In particular, the main steps are: (i)~find
valid inequalities satisfied by any assignment for which $\E[ \norm{
	\mathbf{L} }_q] \leq M$, (ii) reduce the problem to a deterministic
assignment problem for which any feasible solution satisfies the valid
inequalities above, (iii) solve the deterministic problem by writing a
convex programming relaxation, and give a rounding procedure for a
fractional solution to this convex program, and (iv) prove that the
resulting assignment of jobs to machines has small $q$-norm of the load
vector.

\subsection{Useful Bounds}

We start with stating some valid inequalities satisfied by any
assignment $\{J_i\}_{i=1}^m$. 
For each $j\in [n]$ define $Y_j=X_{ij}$ where $j\in J_i$. By definition
of $M$, we know that
\begin{equation}\label{eq:q-norm-bnd}
\E\left[\left(\sum_{i=1}^m (\sum_{j\in J_i} Y_j)^q\right)^{1/q}\right] \le M.
\end{equation}

As in Section~\ref{sec:makespan}, we split each random variable $Y_j$
into two parts: \normal $Y'_j = Y_j\cdot \mathbf{I}_{Y_j\le M}$ and
exceptional $Y''_j = Y_j\cdot \mathbf{I}_{Y_j> M}$. The claim below is
analogous to~(\ref{eq:new-lb-1}), and states that the total expected
size of the exceptional parts cannot be too large.

\begin{claim}\label{cl:norm-LB1}
	For any schedule satisfying~\eqref{eq:q-norm-bnd}, we have
	$\sum_{j=1}^n \E[Y''_j] \le 2M$.
\end{claim}
\begin{pf}
	Suppose for a contradiction that $\sum_{j=1}^n \E[Y''_j] >
	2M$. Lemma~\ref{lem:KRT-max-sum} implies $\E[\max_{j=1}^n Y''_j]
	>M$. Now using the monotonicity of norms and the fact that $Y''_j\le Y_j$, we have
	$$\max_{j=1}^n Y''_j\le \|(Y''_1,\cdots,Y''_n)\|_q\le
	\left(\sum_{i=1}^m (\sum_{j\in J_i} Y_j)^q\right)^{1/q}, $$ which
	contradicts \eqref{eq:q-norm-bnd}.
\end{pf}

Our next two bounds deal with the truncated r.v.s\ $Y_j'$. The first one
states that if we replace $Y_j'$ by its expectation $\E[Y_j']$, the
$q$-norm of this load vector of expectations cannot exceed $M$. The
second bound states that the expected $q^{th}$ moment of the vector
$(Y_j')_{j=1}^n$ is bounded by a constant times $M^q$.
\begin{claim}
	\label{cl:norm-LB2}
	For any schedule satisfying~\eqref{eq:q-norm-bnd} we have
	\[ \sum_{i=1}^m \bigg(\sum_{j\in J_i} \E[Y'_j]\bigg)^q\le M^q. \]
\end{claim}
\begin{pf}
	Since the function $$f(Y'_1,\cdots Y'_n) := \left(\sum_{i=1}^m
	(\sum_{j\in J_i} Y'_j)^q\right)^{1/q}$$ is a norm and hence convex,
	Jensen's inequality (Theorem~\ref{thm:jensen}) implies
	$\E[f(Y'_1,\cdots Y'_n)] \ge f(\E[Y'_1],\cdots \E[Y'_n])$. Raising
	both sides to the $q^{th}$ power and using~\eqref{eq:q-norm-bnd}, the
	claim follows.
\end{pf}

\begin{claim}\label{cl:norm-LB3}
	Let $\alpha=2^{q+1}+8$.  For any schedule
	satisfying~\eqref{eq:q-norm-bnd} we have
	\[ \sum_{j=1}^n \E[(Y'_j)^q] \le \alpha\cdot M^q. \] 
\end{claim}
\begin{pf}
	Define $Z:= \sum_{j=1}^n (Y'_j)^q$ as the quantity of
	interest. Observe that it is the sum of independent $[0,M^q]$ bounded
	random variables. Since $q\ge 1$ and the r.v.s are non-negative, $Z\le
	\sum_{i=1}^m (\sum_{j\in J_i} Y'_j)^q$. Thus \eqref{eq:q-norm-bnd}
	implies $\E[Z^{1/q}] \le M$. However, now Jensen's inequality cannot
	help upper-bound $\E[Z]$.
	
	Instead we use a second-moment calculation. To reduce notation let
	$Z_j :=(Y'_j)^q$, so $Z=\sum_{j=1}^n Z_j$. The variance of $Z$ is
	$var(Z)=\E[Z^2] - \E[Z]^2 \le \sum_{j=1}^n \E[Z_j^2]\le M^q\cdot
	\E[Z]$ as each $Z_j$ is $[0,M^q]$ bounded. By Chebyshev's inequality,
	\begin{align*}
	\Pr\left[Z < \frac{\E[Z]}2 - 4M^q\right] \le \frac{var(Z)}{(\E[Z]/2
		+ 4M^q )^2}
	\le \frac{var(Z)}{(\E[Z] /2)\cdot 4M^q}\le
	\frac{2M^q\cdot E[Z]}{\E[Z]\cdot 4 M^q}\le \frac12.
	\end{align*}
	This implies 
	\begin{gather*}
	E[Z^{1/q}] \ge \frac12 \left( \frac{\E[Z]}2 - 4M^q \right)^{1/q}
	\end{gather*}
	Using the bound $\E[Z^{1/q}] \le M$ from above, we now obtain
	$\E[Z]\le 2\cdot \left( (2M)^q + 4M^q\right)$ as desired.
\end{pf}

In the next sections, we show that the three bounds above are enough to
get a meaningful lower bound on the optimal $q$-norm of load.

\subsection{Reduction to a Deterministic Scheduling Problem}
\label{sec:reduct-determ-sched}

We now formulate a surrogate deterministic scheduling problem, which we
call \qnormsched. An instance of this problem has $n$ jobs and $m$
machines. For each job $j$ and machine $i$, there is a processing time
$p_{ij}$ and two costs $c_{ij}$ and $d_{ij}$. There are also bounds $C$
and $D$ on the two cost functions respectively.  The goal is to find an
assignment of jobs to machines that minimizes the $q$-norm of the
machine loads subject to the constraint that the total $c$-cost and
$d$-cost of the assignments are at most $C$ and $D$ respectively. We now
show how to convert an instance $\I_{stoc}$ of the (stochastic) expected
$q$-norm minimization problem to an instance $\I_{det}$ of the
(deterministic) \qnormsched problem.

Suppose $\I_{stoc}$ has $m$ machines and $n$ jobs, with random variables
$X_{ij}$ for each machine $i$ and job $j$. As before, let $X'_{ij}=
X_{ij}\cdot \mathbf{I}_{X_{ij}\le M}$ and $X''_{ij}=X_{ij}\cdot
\mathbf{I}_{X_{ij}> M}$ denote the truncated and exceptional parts of
each random variable $X_{ij}$ respectively. Then instance $\I_{det}$ has
the same set of jobs and machines as those in $\I$. Furthermore, define
\begin{itemize}
	\item the processing time $p_{ij} = \E[X'_{ij}]$,
	\item the $c$-cost $c_{ij} = \E[X''_{ij}]$ with bound $C=2M$, and 
	\item the $d$-cost $d_{ij} = \E[(X'_{ij})^q]$ with bound $D=\alpha\cdot M^q$. 
\end{itemize}

\begin{observation}
	\label{obs:stoc-det}
	If there is any schedule of expected $q$-norm at most $M$ in the
	instance $\I_{stoc}$, then optimal value of the instance $\I_{det}$ is
	at most $M$.
\end{observation}
\begin{pf}
	This follows directly from Claims~\ref{cl:norm-LB1}, \ref{cl:norm-LB2}
	and \ref{cl:norm-LB3}.
\end{pf}

\subsection{Approximation Algorithm for \qnormsched} 

Our approximation algorithm for the \qnormsched problem is closely based
on the algorithm for unrelated machine scheduling to minimize
$\ell_q$-norms~\cite{AzarE05}. We show:
\begin{theorem} 
	There is a polynomial-time algorithm that given any instance $\I_{det}$ of
	\qnormsched, finds a schedule with (i) $q$-norm of processing times at
	most $2^{1+2/q}\cdot OPT(\I_{det})$, (ii) $c$-cost at most $3C$ and (iii)
	$d$-cost at most $3D$.  \label{thm:det-qnorm}
\end{theorem}
\begin{pf}  We only provide a sketch as many of these
	ideas parallel those from~\cite{AzarE05}. Start with a convex
	programming ``relaxation'' with variables $x_{ij}$ (for assigning job
	$j$ to machine $i$).
	\begin{align*}
	\min  \quad&\sum_{i=1}^m \ell_i^q + \sum_{ij} p_{ij}^q \cdot x_{ij}\\
	\mbox{s.t.}\quad& \ell_i = \sum_j p_{ij}\cdot x_{ij},\, \quad\forall i,\\
	&\sum_i x_{ij} =1,\, \quad\forall j, \\
	&\sum_{ij}c_{ij}\cdot x_{ij} \le C,\\
	&\sum_{ij}d_{ij}\cdot x_{ij} \le D .
	\end{align*}
	 This convex program
	can be solved to arbitrary accuracy and its optimal value is $V\le
	2\cdot OPT(\I_{det})^q$. Let $(x,\ell)$ denote the optimal fractional
	solution below.
	
	We now further reduce this $q$-norm problem to \GAP. The \GAP instance
	$\I_{gap}$ has the same set of jobs and machines as those in $\I_{det}$.  For a
	job $j$ and machine $i$, the processing time remains
	$p_{ij}$. However, the cost of assigning $j$ to $i$ is now
	$\gamma_{ij} := \frac{c_{ij}}{C} + \frac{d_{ij}}{D}
	+\frac{p_{ij}^q}{V}$. Furthermore, we impose a bound of $\ell_i$ on
	the total processing time of jobs assigned to each machine $i$ (i.e.,
	the makespan on $i$ is constrained to be at most $\ell_i$). Note that
	the solution $x$ to the convex program is also a feasible fractional
	solution to the natural LP-relaxation for \GAP with an objective
	function value of $\sum_{ij} \gamma_{ij}\cdot x_{ij}\le 3$. The
	rounding algorithm in \cite{ShmoysT93} can now be used to round $x$
	into an integral assignment $\{A_{ij}\}$ with $\gamma$-cost also at
	most $3$, and load on each machine $i$ being  $L_i\le \ell_i + m_i$,
	where $m_i$ denotes the maximum processing time of any job assigned to
	machine $i$ by this algorithm.  The definition of $\gamma$ and the
	bound on the $\gamma$-cost implies that the $c$-cost and $d$-cost of
	this assignment are at most $3C$ and $3D$ respectively. To bound the
	$q$-norm of processing times,
	\begin{align*}
	\sum_{i=1}^m L_i^q &\le 2^{q-1} \left( \sum_i \ell_i^q + \sum_i
	m_i^q\right)\le 2^{q-1} \left( V + \sum_{ij} p_{ij}^q \cdot
	A_{ij}\right)\le 2^{q-1} (V+3V) = 2^{q+1}\cdot V.
	\end{align*}

	Above, the first inequality uses $(a+b)^q \leq 2^{q-1}(a^q + b^q)$,
	and the third inequality uses the fact that $p_{ij}^q A_{ij} \leq V
	\cdot \gamma_{ij} A_{ij} \leq 3V$ by the bound on the $\gamma$ cost.
	The proof is now completed by using $V\le 2\cdot OPT(\I_{det})^q$.
\end{pf}

\subsection{Interpreting the Rounded Solution}

Starting from an instance $\I_{stoc}$ of expected $q$-norm minimization
problem, we first constructed an instance $\I_{det}$ of \qnormsched.
Let $\mathcal{J}=(J_1,\dots,J_m)$ denote the solution found by applying
Theorem~\ref{thm:det-qnorm} to the instance $\I_{det}$. If the $q$-norm
of processing times of this assignment (as a solution for $\I_{det}$) is
more than $2^{1+2/q} M$ then using Observation~\ref{obs:stoc-det} and
Theorem~\ref{thm:det-qnorm}, we obtain a certificate that the optimal
value of $\I_{stoc}$ is more than $M$. So we assume that ${\cal J}$ has
objective at most $2^{1+2/q} M$ (as a solution to $\I_{det}$). We use exactly 
this assignment as a solution for the stochastic problem as well. It
remains to bound the expected $q$-norm of this assignment.

By the reduction from $\I_{stoc}$ to $\I_{det}$, and the statement of
Theorem~\ref{thm:det-qnorm}, we know that
\begin{align}
\sum_{i=1}^m \sum_{j\in J_i} \E[X''_{ij}] &= \sum_{i=1}^m \sum_{j\in J_i}
c_{ij} \le 6M
\label{eq:norm-ub2} \\
\sum_{i=1}^m \bigg(\sum_{j\in J_i} \E[X'_{ij}]\bigg)^q &= \sum_{i=1}^m (\sum_{j\in
	J_i} p_{ij})^q \le 2^{q+2}\cdot  M^q 
\label{eq:norm-ub1} \\
\sum_{i=1}^m \sum_{j\in J_i}  \E[(X'_{ij})^q]  &= \sum_{i=1}^m \sum_{j\in
	J_i} d_{ij} \le 3\alpha M^q
\label{eq:norm-ub3}
\end{align}
We now derive properties of this assignment as a solution for $\I_{stoc}$. 
\begin{claim}\label{cl:norm-excep-UB}
	The expected $q$-norm of exceptional jobs $\E[(\sum_{i=1}^m
	(\sum_{j\in J_i} X''_{ij})^q)^{1/q}]\le 6M$.
\end{claim}
\begin{pf}
	This follows from~\eqref{eq:norm-ub2}, since the $\ell_q$-norm of a
	vector is at most its $\ell_1$-norm.
\end{pf}

\begin{claim}\label{cl:norm-normal-UB}
	The expected $q$-norm of \normal jobs $\E[(\sum_{i=1}^m (\sum_{j\in
		J_i} X'_{ij})^q)^{1/q}]\le O(\frac{q}{\log q}) M$.
\end{claim}
\begin{pf}
	Define random variables $Q_i := (\sum_{j\in J_i} X'_{ij})^q$, so that
	the $q$-norm of the loads is 
	\[ \textstyle Q := (\sum_{i = 1}^m Q_i )^{1/q}  =
	(\sum_{i=1}^m (\sum_{j\in J_i} X'_{ij})^q)^{1/q}. \] Since $f(Q_1,\cdots Q_m) = (\sum_{i = 1}^m Q_i )^{1/q}$ is a concave function for $q \geq 1$, using Jensen's
	inequality (Theorem~\ref{thm:jensen}) again,
	\begin{equation}\label{eq:norm-calc1}
	\E[Q] \le  \bigg(\sum_{i=1}^m \E[Q_i]\bigg)^{1/q}.
	\end{equation}
	We can bound each $\E[Q_i]$ separately using Rosenthal's inequality (Theorem~\ref{thm:rosen}):
	\begin{align*}
	\E[Q_i] &= \E\bigg[ \big(\sum_{j\in J_i} X'_{ij}\big)^q \bigg]\le
	K^q\cdot \bigg( \big(\sum_{j\in J_i} \E[X'_{ij}]\big)^q + \sum_{j\in
		J_i} \E[(X'_{ij})^q]\bigg),
	\end{align*}
	 where $K=O(q/\log q)$. Summing this
	over all $i = 1, \ldots, m$ and using~\eqref{eq:norm-ub1} and
	\eqref{eq:norm-ub3}, we get
	\begin{equation}
	\label{eq:norm-calc2}
	\sum_{i=1}^m \E[Q_i] \le  K^q \cdot (2^{q+2}+3\alpha)M^q
	\end{equation}
	Recall from Claim~\ref{cl:norm-LB3} that $\alpha = 2^{q+1} + 8$. Now
	plugging this into~\eqref{eq:norm-calc1} we obtain $\E[Q] \le
	O(K)\cdot M$.
\end{pf}

Finally, using Claims~\ref{cl:norm-excep-UB} and \ref{cl:norm-normal-UB}
and the triangle inequality, the expected $q$-norm of solution ${\cal
	J}$ is $O(\frac{q}{\log q})\cdot M$, which completes the proof of
Theorem~\ref{thm:main-qnorm}.
\paragraph{Explicit approximation ratio}
Here we show approximation ratios explicitly. By equations~\eqref{eq:norm-calc1} and~\eqref{eq:norm-calc2}, we have the expected $q$-norm of truncated jobs is 
$$\E[Q]\le (K^q\cdot (2^{q+2}+3\alpha)M^q)^{1/q}=(K^q\cdot (2^{q+2}+3(2^{q+1} + 8))M^q)^{1/q}=(10+3\cdot2^{3-q})^{1/q}2KM.$$
And the expected $q$-norm of exceptional jobs is at most $6M$ by Claim~\ref{cl:norm-excep-UB}. By the triangle inequality, the expected $q$-norm of solution ${\cal J}$ is at most $(6+(10+3\cdot2^{3-q})^{1/q}2K)M.$ Note that for any constant $\epsilon>0$, we can ensure that $M$ is within a $1+\epsilon$ factor of the optimal  value (by the binary search approach). Hence the overall approximation ratio for $q$-norm is $(6+(10+3\cdot2^{3-q})^{1/q}2K)(1+\epsilon)$, for any $\epsilon>0$, where $K$ is the parameter in Theorem~\ref{thm:rosen}. 

The following known result provides a bound of the parameter $K^q$ in Theorem~\ref{thm:rosen}.
\begin{theorem}[\cite{ibragimov2001best}]\label{thm:rosen_cons}
	Let $Z$ denote a random variable with Poisson distribution with parameter 1, i.e., $\Pr[Z=k]=e^{-1}/k!$ for integer $k\ge 0$. Then Theorem~\ref{thm:rosen}  holds with parameter $K = (\E Z^q)^{1/q}$. 
\end{theorem}
\paragraph{Example} For $\ell_2$-norm, $K^2\le \E Z^2=2\Rightarrow K=\sqrt{2}$. The overall approximation ratio is $(6+(10+3\cdot2^{1})^{1/2}2\sqrt{2})(1+\epsilon)=(6+8\sqrt{2})(1+\epsilon)=17.31(1+\epsilon)$. For $\ell_3$-norm, $K^3=5 \Rightarrow K=\sqrt[3]{5}$ and the  approximation ratio   is $(6+(10+3)^{1/3}2\sqrt[3]{5})(1+\epsilon)=14.04(1+\epsilon)$.

\section*{Acknowledgments.}
This work was done in part while the authors were visiting the Simons Institute for the Theory of Computing. A preliminary version of this paper appeared in the proceedings of ACM-SIAM Symposium on Discrete Algorithms 2018. 
Research of A. Gupta is supported in part by NSF awards CCF-1536002, CCF-1540541, and CCF-1617790. V. Nagarajan and X. Shen were supported in part by NSF CAREER grant CCF-1750127.

   \bibliographystyle{alpha}
   \bibliography{bib}
\end{document}